\documentclass[prar,twocolumn,nopacs,amssymb]{revtex4}
\usepackage{graphicx}
\usepackage{amsmath}

\begin{document}

\title{Geometric mass acquisition via quantum metric: \\
an effective band mass theorem for the helicity bands}

\author{M. Iskin}
\affiliation{Department of Physics, Ko\c{c} University, Rumelifeneri Yolu, 
34450 Sar\i yer, Istanbul, Turkey}

\date{\today}

\begin{abstract}

By taking the virtual inter-band processes along with the intra-band ones 
into full account, here we first propose an effective band mass theorem that is
suitable for a wide-class of single-particle Hamiltonians exhibiting multiple 
energy bands. Then, for the special case of two-band systems, we show that 
the inter-band contribution to the effective band mass of a particle at a given 
quantum state is directly controlled by the quantum metric of the corresponding state. 
As an illustration, we consider a spin-orbit coupled spin-$1/2$ particle and 
calculate its effective band mass at the band minimum of the lower helicity band.
Independent of the coupling strength, we find that the bare mass $m_0$ 
of the particle jumps to $2m_0$ for the Rashba and to $3m_0$ for the Weyl coupling. 
This geometric mass enhancement is a non-perturbative effect, uncovering the 
mystery behind the effective mass of the two-body bound states in the non-interacting 
limit. As a further illustration, we show that a massless Dirac quasiparticle 
acquires a linearly dispersing band mass (equivalent to the effective cyclotron 
mass up to a prefactor) with its momentum through the same mechanism.

\end{abstract}


\maketitle

\textit{Introduction.}
Depending on the physical context, the helicity of a particle may refer to 
the projection of its orbital and/or spin angular momentum onto the 
direction of its center of mass motion, e.g., the helicity is said to be right 
(or left) handed when the relevant quantities are aligned (or anti-aligned). 
One of the central themes in a wide-range of modern single-, two-, few- or 
many-body physics problems is the so-called spin-orbit coupling (SOC), 
as it has found a plethora of interdisciplinary applications across atomic 
and molecular, particle, high-energy, nuclear, solid-state and condensed-matter 
physics, spreading over many decades. 
A great deal of these problems involve a
$
\mathbf{d}_\mathbf{k} \cdot \boldsymbol{\sigma}
$
like $\mathbf{k}$-dependent Zeeman term in the equations of motion, where 
the motional degrees of freedom is described by some vector field 
$\mathbf{d}_\mathbf{k}$ that depends on the linear momentum 
$\mathbf{p} = \hbar \mathbf{k}$ through the wave vector $\mathbf{k}$, 
and $\boldsymbol{\sigma}$ is a vector of spin matrices corresponding to the 
spin or generally pseudo-spin (e.g., sublattice, hyperfine, valley, etc.,) 
degrees of freedom. For instance, the generic coupling $\hbar \sum_i \alpha_i k_i \sigma_i$ 
is widespread in physics literature, describing cold atoms in artificial 
non-Abelian gauge fields~\cite{lin11, zhang12, wang12, cheuk12, 
qu13, olson14, jimenez15, wu16, sun17, huang16, meng16}, 
Dirac electrons in low-energy excitations of 
graphene~\cite{novoselov05, zhang05, castro09}, 
and Dirac quasiparticles at the surface of three dimensional topological 
insulators~\cite{hasan10, qi11}, organic quasi-two-dimensional 
materials~\cite{kobayashi07}, and artificial honeycomb like 
(e.g., optical~\cite{tarruell12}, molecular~\cite{gomes12},  
microwave~\cite{bellec13}, etc., ) crystals.

Given the foremost importance of the spin-momentum coupling for the 
general physics community, here we develop a so-called 
\textit{effective band mass theorem} for the helicity bands that is suitable 
for a wide-class of single-particle Hamiltonians.
Our theorem has two physically distinct components: in addition to the usual 
contribution coming from the intra-band processes, there is a geometric 
contribution stemming from the virtual inter-band ones controlled 
by the quantum metric of the corresponding quantum state. 
As an illustration, first we consider a spin-orbit coupled spin-$1/2$ particle, 
and calculate its effective band mass at the band minimum of the lower 
helicity band. It turns out that the bare mass $m_0$ of the particle jumps 
to $2m_0$ for the Rashba and to $3m_0$ for the Weyl coupling 
independently of the coupling strength, i.e., the geometric mass enhancement 
is a non-perturbative effect. Since the quantum metric effects are both rare 
and elusive in nature~\cite{marzari97, zanardi07, haldane11, mudry13, roy15, 
niu14, claassen15, atac15, piechon16, ozawa18, torma15, torma16, iskin18, 
iskin17, iskin18a, iskin18b}, the experimental verification of these predictions 
with the recently realized Rashba like SOCs~\cite{wu16, sun17, huang16, meng16} 
would be an important leap towards the geometric and topological interpretations 
of quantum mechanics~\cite{provost80, berry89, thouless98}.
As a further illustration, then we apply the theorem to a massless Dirac quasiparticle, 
showing that the acquired geometric band mass of the particle disperses linearly 
with its momentum.

It is important to emphasize that the quantum metric contribution to the effective 
band mass of a single particle unveiled in this paper is distinct from our recent 
findings on a counterpart effect on the effective mass of many-body (Cooper pairs) 
as well as two-body (molecular) bound states throughout the BCS-BEC 
crossover in superfluid Fermi gases~\cite{iskin18, iskin18a}. 
These previous works revealed that, through dressing the effective mass of the 
relevant bound state, the quantum metric governs not only the superfluid density 
of some flat-band~\cite{torma15, torma16} and two-band~\cite{iskin18, iskin17} 
Fermi superfluids exhibiting nontrivial quantum geometry, but also many other 
observables including the sound velocity and spin susceptibility~\cite{iskin18b}. 
Unlike all of these previous works on Fermi superfluids where the quantum 
metric is integrated over $\mathbf{k}$ space with some additional weight factors, 
this work paves the direct way towards measuring a local and non-perturbative 
quantum metric effect on either pseudo-spin-$1/2$ Fermi~\cite{huang16, meng16} 
or Bose~\cite{wu16, sun17} gases, independently of the particle statistics.

\textit{Effective band mass theorem for multi-band Hamiltonians.}
Having multiple bands in mind, we consider a generic band structure that is 
determined by the wave equation 
$
H_\mathbf{k} |n \mathbf{k} \rangle = \varepsilon_{n\mathbf{k}}|n \mathbf{k} \rangle,
$
where $H_\mathbf{k}$ is the single-particle Hamiltonian density in reciprocal 
space, and $|n \mathbf{k} \rangle$ and $\varepsilon_{n\mathbf{k}}$ denote, 
respectively, the corresponding energy eigenstates and eigenvalues. 
Here, the quantum states are labeled by the band index $n$ and wave vector 
$\mathbf{k}$. In the presence of such a multi-band setting, the conventional 
wisdom~\cite{am76} for the definitions of the effective band velocity 
$\mathbf{v}_{n\mathbf{k}}$ and effective band mass tensor $\mathbb{M}_{n\mathbf{k}}$ 
of a particle is such that they are determined by the coefficients of the linear 
and quadratic terms in the following small-$\mathbf{q}$ expansion
\begin{align*}
\varepsilon_{n, \mathbf{k}+\mathbf{q}} = \varepsilon_{n \mathbf{k}}
+ \hbar \sum_i v^i_{n\mathbf{k}} q_i 
+ \frac{\hbar^2}{2}\sum_{ij} [\mathbb{M}^{-1}_{n\mathbf{k}}]^{ij} q_i q_j 
+ \cdots.
\end{align*}
Thus, both the components $v^i_{n\mathbf{k}}$ of the velocity and the principal 
values $M^{P}_{n\mathbf{k}}$ of the mass tensor may have strong dependence 
on $\mathbf{k}$ that is controlled by the microscopic details of a given system.

Noting that $\varepsilon_{n, \mathbf{k}+\mathbf{q}}$ is an energy eigenvalue 
of $H_{\mathbf{k}+\mathbf{q}}$, one can simply obtain these effective band 
parameters through perturbation theory. For this purpose, we first perform 
a small-$\mathbf{q}$ expansion of the Hamiltonian
$
H_{\mathbf{k} + \mathbf{q}} = H_\mathbf{k} 
+ \sum_i q_i \partial H_\mathbf{k}/\partial k_i
+ (1/2) \sum_{ij} q_i q_j \partial^2 H_\mathbf{k}/(\partial k_i \partial k_j)
+ \cdots,
$
and then treat all of the $\mathbf{q}$-dependent terms as a perturbative correction
$V_{\mathbf{k} \mathbf{q}}$ to the unperturbed Hamiltonian $H_\mathbf{k}$, i.e.,  
$
H_{\mathbf{k} + \mathbf{q}} = H_\mathbf{k} + V_{\mathbf{k} \mathbf{q}},
$
in the small $\mathbf{q}$ limit. Assuming that $V_{\mathbf{k} \mathbf{q}}$ 
connects $|n \mathbf{k} \rangle$ state only to non-degenerate ones, this 
approach gives
$
\varepsilon_{n, \mathbf{k}+\mathbf{q}} = \varepsilon_{n\mathbf{k}}
+ \langle n \mathbf{k} | V_{\mathbf{k} \mathbf{q}} | n \mathbf{k} \rangle
+ \sum^\prime 
|\langle n \mathbf{k} | V_{\mathbf{k} \mathbf{q}} | n' \mathbf{k'} \rangle|^2
/(\varepsilon_{n\mathbf{k}} - \varepsilon_{n'\mathbf{k'}})
+ \cdots.
$
In principle, 
$
\sum^\prime \equiv \sum_{\{n'\mathbf{k'}\} \ne \{n\mathbf{k}\}}
$ 
sums over all of the non-degenerate quantum states, but it is sufficient to keep
only the inter-band terms with $\mathbf{k'} = \mathbf{k}$ in the small $\mathbf{q}$ limit.
By matching the coefficients of the linear and quadratic terms in $\mathbf{q}$, 
we eventually identify
$
\hbar v^i_{n\mathbf{k}} = 
\langle n\mathbf{k} | \partial H_\mathbf{k}/\partial k_i | n\mathbf{k} \rangle
= \partial \varepsilon_{n \mathbf{k}}/ \partial k_i
$
as the effective band velocity, and 
\begin{align}
\label{eqn:emt}
[\mathbb{M}^{-1}_{n\mathbf{k}}]^{ij}
&= \frac{1}{\hbar^2}\left\langle n\mathbf{k} \bigg | \frac{\partial^2 H_\mathbf{k}}{\partial k_i \partial k_j}
 \bigg | n\mathbf{k} \right \rangle \\
&+
2\mathrm{Re}\sum^\prime  \frac{
\left\langle n\mathbf{k} \big | \frac{\partial H_\mathbf{k}}{\partial k_i}
 \big | n' \mathbf{k} \right \rangle
\left\langle n'\mathbf{k} \big | \frac{\partial H_\mathbf{k}}{\partial k_j}
 \big | n\mathbf{k} \right \rangle}
{\hbar^2(\varepsilon_{n\mathbf{k}} - \varepsilon_{n'\mathbf{k}})} \nonumber
\end{align}
as the inverse effective band mass tensor of the particle for the $\mathbf{k}$ state. 
Here, $\mathrm{Re}$ denotes the real part of the resultant summation. 
In Eq.~(\ref{eqn:emt}), while the first term on the right is due precisely to the 
intra-band contribution 
$
[\textbf{\textrm{m}}^{-1}_{n\mathbf{k}}]^{ij}
$
to the effective inverse mass tensor of the corresponding energy band, 
the second term is generated by the $\mathbf{q}$-induced virtual 
inter-band processes.

Despite the use of perturbation theory in its derivation, Eq.~(\ref{eqn:emt}) is 
exact, and in principle, the inter-band contribution need not necessarily be a 
small correction to the intra-band one. For instance, one can alternatively 
obtain Eq.~(\ref{eqn:emt}) by taking the derivatives of the wave equation
$
H_\mathbf{K} | n\mathbf{K} \rangle = \varepsilon_{n\mathbf{K}} |n\mathbf{K} \rangle
$
with respect to $\mathbf{K} = \mathbf{k} + \mathbf{q}$ as follows.
Using the orthonormalization condition 
$
\langle n\mathbf{K} | n'\mathbf{K'} \rangle = \delta_{nn'} \delta(\mathbf{K}-\mathbf{K'})
$
for the $\mathbf{K}$ subspace of energy eigenstates, where $\delta_{nn'}$ is a 
Kronecker-delta and $\delta(x)$ is a Dirac-delta function, we first obtain
$
\langle n\mathbf{K} | \partial H_\mathbf{K}/\partial K_i |n' \mathbf{K'} \rangle = 
(\varepsilon_{n\mathbf{K}} - \varepsilon_{n'\mathbf{K'}}) 
(\partial \langle n\mathbf{K} |/\partial K_i) | n'\mathbf{K'} \rangle 
+ (\partial \varepsilon_{n\mathbf{K}}/\partial K_i) \delta_{nn'} \delta(\mathbf{K}-\mathbf{K'}),
$
leading to
$
\partial \varepsilon_{n\mathbf{K}}/\partial K_i = 
\langle n\mathbf{K} | \partial H_\mathbf{K}/\partial K_i |n \mathbf{K} \rangle.
$
Then, by taking the derivative of this expression with respect to $K_j$, i.e., 
$
\partial^2 \varepsilon_{n\mathbf{K}}/(\partial K_i \partial K_j) = 
\langle n\mathbf{K} | \partial^2 H_\mathbf{K}/(\partial K_i \partial K_j) |n \mathbf{K} \rangle 
+ (\partial \langle n\mathbf{K} |/\partial K_j) |  \partial H_\mathbf{K}/\partial K_i | n \mathbf{K} \rangle
+ \langle n\mathbf{K} | \partial H_\mathbf{K}/\partial K_i | (\partial | n\mathbf{K} \rangle /\partial K_j),
$
and using the completeness relation
$
\mathbb{I} = \sum_{n\mathbf{K}} | n\mathbf{K} \rangle \langle n\mathbf{K} |
$
together with the derivative of the orthonormalization condition
$
(\partial \langle n\mathbf{K} |/\partial K_i) | n'\mathbf{K} \rangle 
+ \langle n\mathbf{K} | (\partial | n'\mathbf{K} \rangle / \partial K_i) = 0
$
for the $\mathbf{K}$ subspace of interest, we eventually arrive at
\begin{align}
\label{eqn:emta}
\frac{\partial^2 \varepsilon_{n\mathbf{K}}}{\partial K_i \partial K_j} 
&= 
\left\langle n\mathbf{K} \bigg | \frac{\partial^2 H_\mathbf{K}}{\partial K_i \partial K_j}
\bigg | n\mathbf{K} \right \rangle \\
+ 2 \mathrm{Re} & \sum^\prime
(\varepsilon_{n\mathbf{K}} - \varepsilon_{n'\mathbf{K'}}) 
\frac{\partial \langle n\mathbf{K} |}{\partial K_i} | n'\mathbf{K'} \rangle 
\langle n'\mathbf{K'} | \frac{\partial | n\mathbf{K} \rangle}{\partial K_j} \nonumber.
\end{align}
This expression corresponds precisely to the coefficient 
$\hbar^2 [\mathbb{M}^{-1}_{n\mathbf{k}}]^{ij}$ of the quadratic term when 
$\mathbf{q} \to \mathbf{0}$, producing Eq.~(\ref{eqn:emt}) without the use 
of perturbative approach.

Having established the theoretical framework on the effective band mass 
theorem for multi-band Hamiltonians, next we focus on two-band systems 
for their simplicity, and highlight not only the intriguing interpretation of the 
inter-band contribution from the quantum geometrical perspective but also 
illustrate its relative importance for a number of toy models that are of immediate 
experimental and/or theoretical interest.

\textit{Quantum metric of the projected Hilbert space.}
For this purpose, we note that the inter-band contribution to the effective band 
mass theorem given in Eq.~(\ref{eqn:emt}) resembles closely, but not quite right, 
to the quantum metric $g_{n \mathbf{k}}^{ij}$ of the corresponding energy 
eigenstate. Recall that~\cite{provost80, berry89, thouless98} 
$
g_{n\mathbf{k}}^{ij} = \mathrm{Re}
(\partial \langle n\mathbf{k}|/\partial k_i) 
(\sum^\prime | n'\mathbf{k'} \rangle \langle n'\mathbf{k'} |)
(\partial |n\mathbf{k} \rangle/\partial k_j)
$
is defined as the real part of the so-called quantum geometric tensor
$
Q_{n\mathbf{k}}^{ij} = g_{n \mathbf{k}}^{ij} - (\mathrm{i}/2) F_{n \mathbf{k}}^{ij}
$
of the projected Hilbert space, which can be equivalently expressed as
$
g_{n\mathbf{k}}^{ij} = \sum_\mathbf{k'} g_{n\mathbf{k \mathbf{k'}}}^{ij}
$
where
\begin{align}
g_{n\mathbf{k}\mathbf{k'}}^{ij} = \mathrm{Re}\sum^\prime \frac{
\left\langle n\mathbf{k} \big | \frac{\partial H_\mathbf{k}}{\partial k_i}
\big | n' \mathbf{k'} \right \rangle
\left\langle n'\mathbf{k'} \big | \frac{\partial H_\mathbf{k}}{\partial k_j}
\big | n\mathbf{k} \right \rangle}
{(\varepsilon_{n\mathbf{k}} - \varepsilon_{n'\mathbf{k'}})^2}.
\label{eqn:qm}
\end{align}
Since only the inter-band terms with $\mathbf{k'} = \mathbf{k}$ survive 
in Eqs.~(\ref{eqn:emt}) or~(\ref{eqn:emta}), we have
$
g_{n\mathbf{k}}^{ij} \equiv g_{n\mathbf{k \mathbf{k}}}^{ij}
$
in this paper. 
The geometrical importance of the quantum metric reveals itself as a measure 
of the quantum distance between differing quantum states, i.e.,
$
ds_n^2 =  \langle n\mathbf{k} | n \mathbf{k} \rangle 
- |\langle n \mathbf{k} | n, \mathbf{k}+d\mathbf{k} \rangle|^2 
= \sum_{\mu\nu} g_{n \mathbf{k}}^{ij} dk_i dk_j.
$
In contrast, the imaginary part $F_{n \mathbf{k}}^{ij}$ of the quantum geometric 
tensor is the Berry curvature, i.e., it is a distinct but related quantity corresponding 
to the emergent magnetic field in $\mathbf{k}$ space. 

Even though both $g_{n \mathbf{k}}^{ij}$ and $F_{n \mathbf{k}}^{ij}$ characterize 
the local $\mathbf{k}$-space geometry of the underlying quantum states, 
they may also be linked with the global properties of the system in somewhat 
peculiar ways. For instance, the topological Chern invariant of a quantum-Hall 
system is determined by an integration of $F_{n \mathbf{k}}^{ij}$ in $\mathbf{k}$ 
space, i.e., 
$
2\pi C_n = \int_{BZ} dk_x dk_y F_{n \mathbf{k}}^{xy}
$
controls the Hall conductivity~\cite{thouless98}. Likewise, the superfluid density 
of some flat-band and two-band superfluids, that exhibit nontrivial quantum geometry, 
has a substantial contribution determined by an integration of $g_{n \mathbf{k}}^{ij}$ 
with a proper weight factor~\cite{torma15, torma16, iskin18, iskin17}. 
Furthermore, by showing that the quantum metric contribution accounts for a 
sizeable fraction of the (Cooper) pair mass throughout the BCS-BEC 
crossover~\cite{iskin18, iskin18a}, we have revealed the physical origin of its 
governing role in the superfluid density and hinted at its plausible roles in many 
other observables including the sound velocity and spin susceptibility~\cite{iskin18b}. 
Next we point out a distinct but related effect on the effective band mass 
of a single particle nearby the band minimum of the lower helicity band.

\textit{Effective band mass theorem for two-band Hamiltonians.}
We consider a generic two-band Hamiltonian density $H_\mathbf{k}$, 
and parametrize its energy eigenstates and eigenvalues 
with $|s \mathbf{k} \rangle$ and
$
\varepsilon_{s\mathbf{k}} = \epsilon_\mathbf{k} + s d_\mathbf{k},
$
respectively. Here, $s = \pm$ labels the upper/lower bands that are separated
by a $\mathbf{k}$-dependent energy gap of $2d_\mathbf{k}$.
When the geometric response is due only to the intermediate states from the 
differing band, i.e., $V_{\mathbf{k}\mathbf{q}}$ connects $|s \mathbf{k} \rangle$ 
to $|-s, \mathbf{k} \rangle$ with $d_\mathbf{k} \ne 0$, a compact way to express 
the resultant effective band mass theorem for such a two-band system is
\begin{align}
\label{eqn:emt2}
[\mathbb{M}^{-1}_{s\mathbf{k}}]^{ij} = [\textbf{\textrm{m}}^{-1}_{s\mathbf{k}}]^{ij} 
+ \frac{2s}{\hbar^2} d_\mathbf{k} g_{\mathbf{k}}^{ij}.
\end{align}
Here, the intra-band contribution is simply denoted as 
$[\textbf{\textrm{m}}^{-1}_{s\mathbf{k}}]^{ij}$,
but the inter-band one is directly controlled by the total quantum metric 
$
g_{\mathbf{k}}^{ij} = \sum_s g_{s\mathbf{k}}^{ij} = \mathrm{Re}
\sum_s (\partial \langle s\mathbf{k}|/\partial k_i) 
(|-s,\mathbf{k} \rangle \langle -s,\mathbf{k} |)
(\partial | s\mathbf{k} \rangle/\partial k_j)
$
of the corresponding quantum state. Note that the inter-band contributions 
have opposite signs for the $s=\pm$ bands. For the isotropic systems of
interest in this paper, the relevant effective mass is defined as the 
$\mathbf{k}$-space average of Eq.~(\ref{eqn:emt2}) over the degenerate 
subspace.

To illustrate the relative importance of the inter-band contribution, next we 
consider a single spin-$1/2$ particle that is described by the generic 
Hamiltonian density
$
H_\mathbf{k} = \epsilon_{\mathbf{k}} \sigma_0 + \mathbf{d}_\mathbf{k} \cdot \boldsymbol{\sigma},
$
where $\epsilon_k = \hbar^2 k^2/(2m_0)$ is the usual free-particle 
dispersion with $m_0$ the bare mass, and
$
\mathbf{d}_\mathbf{k} = \sum_i d_\mathbf{k}^i \boldsymbol{\widehat{i}}
$
is the SOC field with $\boldsymbol{\widehat{i}}$ the unit vector along 
the $i$ direction. Here, $\sigma_0$ is a $2 \times 2$ identity matrix, and 
$
\boldsymbol{\sigma} = \sum_i \sigma_i \boldsymbol{\widehat{i}}
$
is a vector of Pauli spin matrices in such a way that
$
d_\mathbf{k}^i = \hbar \alpha_i k_i
$
corresponds to the Weyl SOC when $\alpha_i = \alpha$ for all $i = \{x,y,z\}$,
a Rashba SOC when $\alpha_z = 0$, and to an equal Rashba-Dresselhaus
(ERD) SOC when $\alpha_y = \alpha_z = 0$. Here, we choose $\alpha \ge 0$ 
without the loss of generality. Note that 
$
\varepsilon_{s\mathbf{k}} = \epsilon_\mathbf{k} + s d_\mathbf{k}
$
is the $s$-helicity dispersion with the following energy eigenstate
$
|s \mathbf{k} \rangle^\textrm{T} = (-d_\mathbf{k}^x+\mathrm{i} d_\mathbf{k}^y, 
d_\mathbf{k}^z - s d_\mathbf{k})/\sqrt{2d_\mathbf{k}(d_\mathbf{k} - s d_\mathbf{k}^z)},
$
where $d_\mathbf{k} = |\mathbf{d}_\mathbf{k}|$ and $\textrm{T}$ is the transpose 
operator. The quantum geometry is trivial for the ERD SOC as 
$
g_{s\mathbf{k}}^{ij} = 0
$ 
for the entire $\mathbf{k}$ space.
It turns out that the quantum metrics of the $s = \pm$ bands are equal, and their
sum can be expressed as
$
g_{\mathbf{k}}^{ij} = (\partial \boldsymbol{\widehat{d}_k} \partial k_i)
\cdot (\partial \boldsymbol{\widehat{d}_k} \partial k_j)/2,
$
where
$
\boldsymbol{\widehat{d}_k} = \mathbf{d}_\mathbf{k}/d_\mathbf{k}
$
is the unit vector along the SOC field. The equivalent expression
$
g_{\mathbf{k}}^{ij} = [\sum_\ell (\partial d_\mathbf{k}^\ell/\partial k_i) 
(\partial d_\mathbf{k}^\ell/\partial k_j) 
- (\partial d_\mathbf{k}/\partial k_i) (\partial d_\mathbf{k}/\partial k_j)]/(2d_\mathbf{k}^2)
$
reduces simply to 
$
g_{\mathbf{k}}^{ij} = \hbar^2 \alpha_i \alpha_j 
(d_\mathbf{k}^2 \delta_{ij} - d_\mathbf{k}^i d_\mathbf{k}^j)
/(2d_\mathbf{k}^4),
$ 
showing that 
$
g_{\mathbf{k}}^{ij} = 0
$
in general for all $\mathbf{k}$ states unless $\alpha_i \alpha_j \ne 0$.

\begin{figure}[htbp]
\includegraphics[scale=0.5]{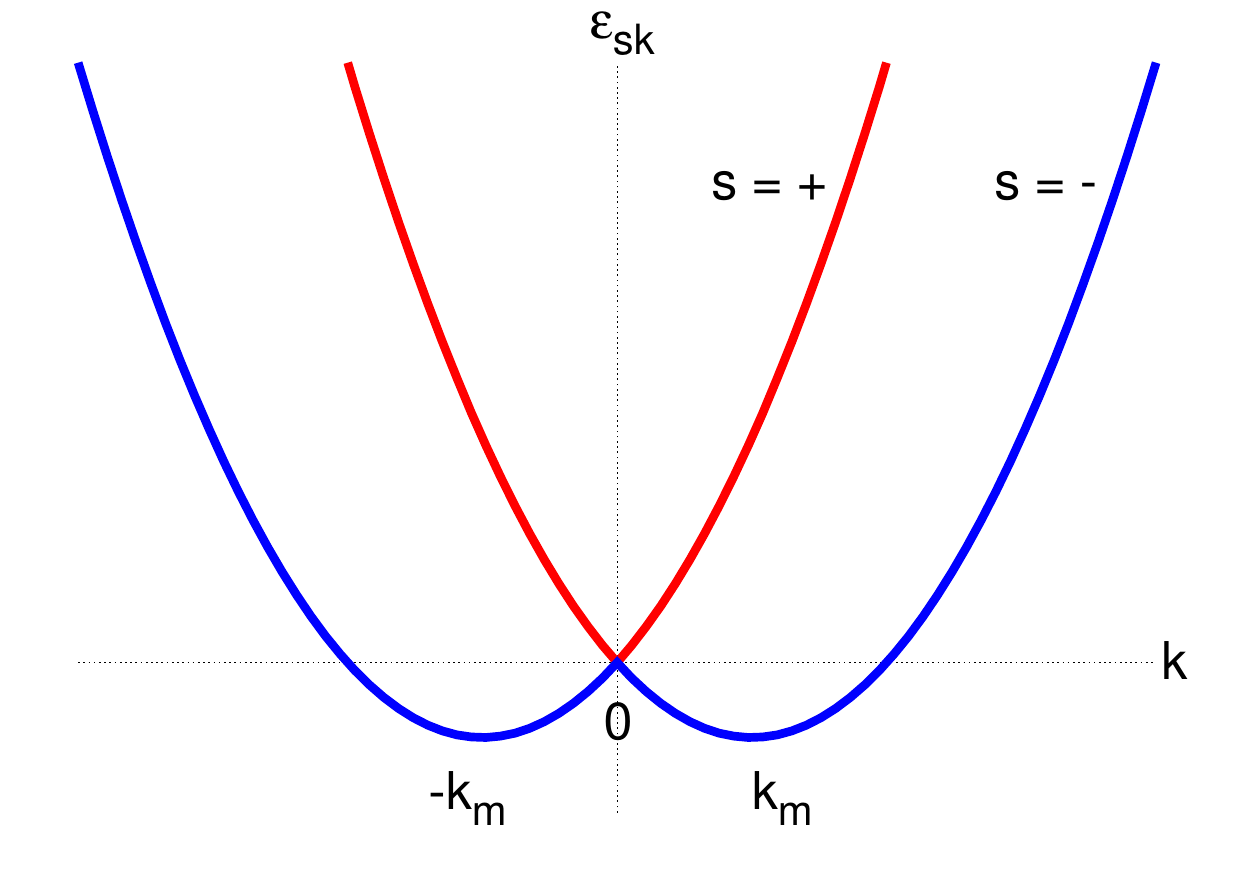}
\caption{(color online)
\label{fig:bands}
By breaking the spin degeneracy of the particle dispersions
$
\epsilon_{\sigma \mathbf{k}} = \epsilon_\mathbf{k},
$ 
the SOC $\mathbf{d}_\mathbf{k} \cdot \boldsymbol{\sigma}$ creates a pair 
of helicity bands 
$
\varepsilon_{s \mathbf{k}} = \epsilon_\mathbf{k} + sd_\mathbf{k}.
$ 
Here, $s = \pm$ corresponds to the upper/lower branches, and the figure 
illustrates them for a free particle $\epsilon_k = \hbar^2 k^2/(2m_0)$ 
with Weyl SOC $d_k = \hbar \alpha k$.
We are interested in the effective band mass of the particle nearby the 
band minimum ($\mathbf{k_m}$) of the lower helicity band.
}
\end{figure}

In Eq.~(\ref{eqn:emt2}), the intra-band contribution $m_{sk} = m_0$ is already 
isotropic in $\mathbf{k}$ space, and we take the angular average of the 
geometric contribution over the degenerate direction, i.e.,
$
\langle g_{\mathbf{k}}^{ij} \rangle = g_{k} \delta_{ij}
$
with
$
g_{k} = \hbar^2 \alpha^2 (D-1) / (2D d_k^2).
$
Here, $D = 3$ for the radial direction of Weyl SOC with $d_k = \hbar \alpha k$, 
$D = 2$ for the polar direction of Rashba SOC with $d_k = \hbar \alpha k_\perp$, 
and $D = 1$ for the $q_x$ direction of ERD SOC with $d_k = \hbar \alpha |k_x|$. 
This leads to
\begin{equation}
\label{eqn:mk}
\frac{M_{sk}}{m_0} = \frac{D d_k}{D d_k + s(D-1)m_0 \alpha^2},
\end{equation}
for the angular average 
$
\langle [\mathbb{M}^{-1}_{s\mathbf{k}}]^{ij} \rangle = \delta_{ij}/M_{sk}
$
of the effective mass theorem, showing that the geometric contribution is 
negligible at large momentum, i.e., when $D d_k \gg (D-1) m_0\alpha^2$.
In this paper, we are interested in the effective band mass of the particle nearby 
the band minimum $\mathbf{k_m}$ of the lower helicity band, e.g., the Weyl SOC 
is illustrated in Fig.~\ref{fig:bands}. Plugging $k_m = m_0\alpha/\hbar$ in 
Eq.~(\ref{eqn:mk}), we find
\begin{equation}
M_{-,k_m} = D m_0,
\label{eqn:mkm}
\end{equation}
where $D$ is the dimensionality of the isotropic SOC field. 

\begin{figure}[htbp]
\includegraphics[scale=0.6]{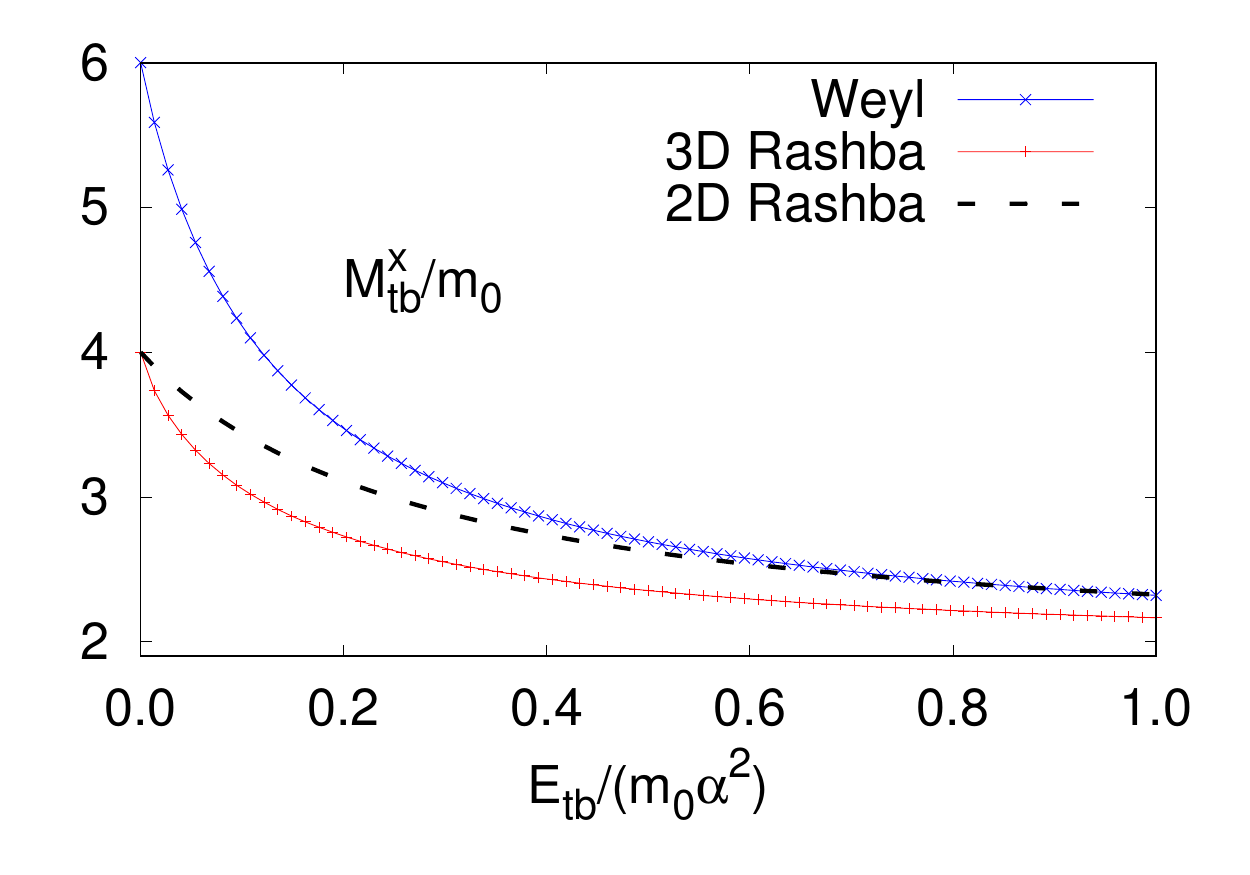}
\caption{(color online)
\label{fig:pairmass}
The effective mass of the two-body bound state $M_{tb}^{x}$ is shown for the 
Weyl and Rashba SOCs as a function of the two-body binding energy $E_{tb}$ in 
vacuum~\cite{zhai11, iskin11, jiang11, hu11, he12a, he12b, shenoy12a, shenoy12b, torma18}. 
In the non-interacting limit when $E_{tb} \to 0^+$, we note that $M_{tb}^{x}$ 
approaches twice the effective band mass of the particle at the band minimum 
of its lower helicity band, i.e., $M_{tb} \to 2M_{-,k_m}$ in all cases 
considered here.
}
\end{figure}

This analysis implies that the inter-band contribution to the effective band mass 
of the particle is a non-perturbative one for both Weyl and Rashba SOCs.
In addition, this contribution is clearly independent of $\alpha$ 
at the band minimum as long as $\alpha \ne 0$. Furthermore, the dimensionality
of the $\mathbf{k}$ space does not play any role for the Rashba SOC, i.e., 
$M_{-,k_m}^{\perp} = 2m_0$ in both two and three dimensions.
No matter how bizarre these results sound, we note that they are consistent 
with the effective mass $M_{tb}$ of the relevant two-body bound states 
reported in the Fermi gas 
literature~\cite{zhai11, iskin11, jiang11, hu11, he12a, he12b, shenoy12a, shenoy12b, torma18},
where the binding occurs between an $(\uparrow; \mathbf{k}+\mathbf{q}/2)$ 
particle and a $(\downarrow; -\mathbf{k}+\mathbf{q}/2)$ one in the 
presence of a contact attraction. For instance, in Fig.~\ref{fig:pairmass}, we 
reproduce $M_{tb}^{x}$ as a function of the two-body binding energy 
$E_{tb}$ in vacuum. 
In the non-interacting limit when $E_{tb} \to 0^+$, we clearly see that 
$M_{tb} = 6m_0$ for the Weyl SOC and $M_{tb}^{\perp} = 4m_0$ for the 
Rashba SOC as long as $\alpha \ne 0$, independently of its magnitude. 
In addition, it is also known that $M_{tb}^{z} = 2m_0$ for the Rashba SOC in 
three dimensions, and that the ERD SOC has no effect on the two-body problem.
Thus, as $M_{tb}$ coincides with twice the effective band mass of the particle 
at the band minimum of its lower helicity band, the geometric mass enhancement 
uncovers the mystery behind $M_{tb}$ in the non-interacting limit. 
When $E_{tb}$ increases from $0$, however, a competing but distinct geometric 
effect on the bound state reduces $M_{tb}$ towards the expected value 
$2m_0$ in the $E_{tb} \gg m_0 \alpha^2$ limit~\cite{iskin18a}.

We note in passing that the electronic properties of some graphene-like 
solid-state materials are well-described by the low-energy Hamiltonian 
density
$
H_\mathbf{k} = \hbar \alpha \mathbf{k} \cdot \boldsymbol{\sigma},
$
where $\alpha$ refers to the Fermi velocity, and $\boldsymbol{\sigma}$ refers 
not to the spin degrees of freedom of the electron but to the sub-lattice 
degrees of freedom of the underlying crystal (i.e., honeycomb) lattice~\cite{castro09}. 
Since the helicity dispersions 
$
\varepsilon_{sk} = s \alpha k
$
are isotropic and linear in $\mathbf{k}$ space, the intra-band contribution to the
inverse effective mass vanishes, suggesting that the effective band mass 
$M_{sk}$ is determined solely by the inter-band contribution. Thus, setting 
$m_0 \to \infty$ in Eq.~(\ref{eqn:mk}) we find a linearly dispersing effective 
band mass 
\begin{equation}
M_{sk} = \frac{sD}{D-1} \frac{\hbar k}{\alpha}
\end{equation}
for the particles/holes in the upper/lower helicity bands, where $D \ge 2$ 
is the dimensionality of the $\mathbf{k}$ space.
Indeed, the long-established realization that the effective cyclotron mass
$
M_{sc} = s\hbar k_F/\alpha 
$
of the charge carriers in graphene has a strong $\sqrt{\rho}$ dependence 
in the low-density limit, provided one of the earliest smoking-gun evidences 
for the existence of a massless Dirac quasiparticle in the $k \to 0^+$ 
limit~\cite{novoselov05, zhang05, castro09}.
Here, $\rho = k_F^2/\pi$ is the electronic density of graphene with $k_F$ 
the Fermi wave vector.

\textit{Conclusions.}
In summary, our so-called effective band mass theorem for the helicity bands 
has two physically distinct components: in addition to the usual contribution 
coming from the intra-band processes, there exists a geometric contribution 
stemming from the virtual inter-band ones. It turns out that the latter contribution 
is directly controlled by the quantum metric of the corresponding quantum state. 
To illustrate the relative importance of the inter-band contribution, we considered 
a spin-orbit coupled spin-$1/2$ particle, and calculated its effective band mass at 
the band minimum of the lower helicity band. We found that the bare mass $m_0$ 
of the particle jumps to $2m_0$ for the Rashba and to $3m_0$ for the Weyl coupling, 
independently of the coupling strength. We argued that such a geometric 
enhancement of the effective mass of the particle is consistent with 
the effective mass of the relevant two-body bound state in the non-interacting 
limit~\cite{zhai11, iskin11, jiang11, hu11, he12a, he12b, shenoy12a, shenoy12b, torma18}.
In addition, for graphene-like solid-state materials, we noted that the physical 
mechanism behind the linearly dispersing effective cyclotron mass of a Dirac
quasiparticle~\cite{novoselov05, zhang05, castro09} may be interpreted as 
a geometric mass acquisition.

\begin{acknowledgments}
The author thanks F. M. Ramazano\u{g}lu and M. \"O. Oktel for the critical 
reading of the manuscript, and acknowledges support from 
T{\"U}B{\.I}TAK and the BAGEP award of the Turkish Science Academy.
\end{acknowledgments}


\begin{thebibliography}{99}

\bibitem{lin11} Y.-J. Lin, K. Jim\'{e}nez-Garc\'{\i}a, and I. B. Spielman,
``Spin-orbit-coupled BoseÐEinstein condensates", 
Nature \textbf{471}, 83 (2011).

\bibitem{zhang12} J. Y. Zhang, S. C. Ji, Z. Chen, L. Zhang, Z. D. Du,Bo Yan, G. S. Pan, B. Zhao, Y. J. Deng, H. Zhai, S. Chen, and J. W. Pan,
``Collective Dipole Oscillations of a Spin-Orbit Coupled Bose-Einstein Condensate", 
Phys. Rev. Lett. \textbf{109}, 115301 (2012).

\bibitem{wang12} P. Wang, Z.-Q. Yu, Z. Fu, J. Miao, L. Huang, S. Chai, H. Zhai, and J. Zhang, 
``Spin-orbit coupled degenerate Fermi gases", 
Phys. Rev. Lett. \textbf{109}, 095301 (2012).

\bibitem{cheuk12} L. W. Cheuk, A.T. Sommer, Z. Hadzibabic, T. Yefsah, W. S. Bakr, and M.W. Zwierlein, 
``Spin-Injection Spectroscopy of a Spin-Orbit Coupled Fermi Gas", 
Phys. Rev. Lett. \textbf{109}, 095302 (2012).

\bibitem{qu13} C. Qu, C. Hamner, M. Gong, C. Zhang, and P. Engels,
``Observation of Zitterbewegung in a spin-orbit-coupled Bose-Einstein condensate", 
Phys. Rev. A \textbf{88}, 021604(R) (2013).

\bibitem{olson14} A. J. Olson, S.-J. Wang, R. J. Niffenegger, C.-H. Li, C. H. Greene, and Y. P. Chen, 
``Tunable Landau-Zener transitions in a spin-orbit-coupled Bose-Einstein condensate", 
Phys. Rev. A \textbf{90}, 013616 (2014).

\bibitem{jimenez15} K. Jim\'{e}nez-Garc\'{\i}a, L. LeBlanc, R.Williams, M. Beeler, C. Qu, M. Gong, C. Zhang, and I. Spielman,
``Tunable Spin-Orbit Coupling via Strong Driving in Ultracold-Atom Systems", 
Phys. Rev. Lett. \textbf{114}, 125301 (2015).


\bibitem{huang16}
L. Huang, Z. Meng, P. Wang, P. Peng, S.-L. Zhang, L. Chen, D. Li, Q. Zhou, and J. Zhang,
``Experimental realization of a two-dimensional synthetic spin-orbit coupling in ultracold Fermi gases",
Nat. Phys. \textbf{12}, 540 (2016).

\bibitem{meng16}
Z. Meng, L. Huang, P. Peng, D. Li, L. Chen, Y. Xu, C. Zhang, P. Wang, and J. Zhang,
``Experimental Observation of a Topological Band Gap Opening in Ultracold Fermi Gases with Two-Dimensional Spin-Orbit Coupling",
Phys. Rev. Lett. \textbf{117}, 235304 (2016).

\bibitem{wu16}
Z. Wu, L. Zhang, W. Sun, X.-T. Xu, B.-Z. Wang, S.-C. Ji, Y. Deng, S. Chen, X.-J. Liu, and J.-W. Pan,
``Realization of Two-Dimensional Spin-orbit Coupling for Bose-Einstein Condensates",
Science \textbf{354}, 83 (2016).

\bibitem{sun17}
W. Sun, B.-Z. Wang, X.-T. Xu, C.-R. Yi, L. Zhang, Z. Wu, Y. Deng, X.-J. Liu, S. Chen, and J.-W. Pan,
``Long-lived 2D Spin-Orbit coupled Topological Bose Gas",
arXiv:1710.00717.


\bibitem{novoselov05}
K. S. Novoselov, A. K. Geim, S. V. Morozov, D. Jiang, M. I. Katsnelson, I. V. Grigorieva, S. V.
Dubonos, and A. A. Firsov,
``Two-dimensional gas of massless Dirac fermions in graphene",
Nature \textbf{438}, 197 (2005).

\bibitem{zhang05}
Y. Zhang, Y.-W. Tan, H. L. Stormer, and P. Kim,
``Experimental observation of the quantum Hall effect and Berry's phase in graphene",
Nature \textbf{438}, 201 (2005).

\bibitem{castro09}
A. H. Castro Neto, F. Guinea, N. M. R. Peres, K. S. Novoselov, and A. K. Geim,
``The electronic properties of graphene",
Rev. Mod. Phys. \textbf{81}, 109 (2009).

 
\bibitem{hasan10} 
M. Z. Hasan and C. L. Kane, 
``Colloquium: Topological insulators", 
Rev. Mod. Phys. \textbf{82}, 3045 (2010).

\bibitem{qi11} X.-L. Qi and S.-C. Zhang, 
``Topological insulators and superconductors", 
Rev. Mod. Phys. \textbf{83}, 1057 (2011).


\bibitem{kobayashi07}
A. Kobayashi, S. Katayama, Y. Suzumura, and H. Fukuyama,
``Massless Fermions in Organic Conductor",
J. Phys. Soc. Jpn. \textbf{76}, 034711 (2007).

\bibitem{tarruell12}
L. Tarruell, D. Greif, T. Uehlinger, G. Jotzu, and Tilman Esslinger,
``Creating, moving and merging Dirac points with a Fermi gas in a tunable honeycomb lattice",
Nature \textbf{483}, 302 (2012).
 
\bibitem{gomes12}
K. K. Gomes, W. Mar, W. Ko, F. Guinea, and H. C. Manoharan,
``Designer Dirac fermions and topological phases in molecular graphene",
Nature \textbf{483},  306 (2012).    

\bibitem{bellec13}
M. Bellec, U. Kuhl, G. Montambaux, and F. Mortessagne,
``Topological Transition of Dirac Points in a Microwave Experiment",
Phys. Rev. Lett. \textbf{110}, 033902 (2013).


\bibitem{marzari97}
N. Marzari and D. Vanderbilt,
``Maximally localized generalized Wannier functions for composite energy bands",
Phys. Rev. B \textbf{56}, 12847 (1997).

\bibitem{zanardi07}
P. Zanardi, P. Giorda, and M. Cozzini,
``Information-Theoretic Differential Geometry of Quantum Phase Transitions",
Phys. Rev. Lett. \textbf{99}, 100603 (2007).

\bibitem{haldane11}
F. D. M. Haldane,
``Geometrical Description of the Fractional Quantum Hall Effect",
Phys. Rev. Lett. \textbf{107}, 116801 (2011).

\bibitem{mudry13}
T. Neupert, C. Chamon, and C. Mudry,
``Measuring the quantum geometry of Bloch bands with current noise",
Phys. Rev. B \textbf{87}, 245103 (2013).

\bibitem{roy15}
T. S. Jackson, G. Moller, and R. Roy, 
``Geometric stability of topological lattice phases", 
Nat. Commun. \textbf{6}, 8629 (2015).

\bibitem{niu14}
Y. Gao, S. A. Yang, and Q. Niu,
``Field Induced Positional Shift of Bloch Electrons and Its Dynamical Implications",
Phys. Rev. Lett. \textbf{112}, 166601 (2014).

\bibitem{claassen15}
M. Claassen, C. H. Lee, R. Thomale, X.-L. Qi, and T. P. Devereaux,
``Position-Momentum Duality and Fractional Quantum Hall Effect in Chern Insulators",
Phys. Rev. Lett. \textbf{114}, 236802 (2015).

\bibitem{atac15}
A. Srivastava and A. Imamoglu,
``Signatures of Bloch-Band Geometry on Excitons: Nonhydrogenic Spectra in Transition-Metal Dichalcogenides",
Phys. Rev. Lett. \textbf{115}, 166802 (2015).

\bibitem{piechon16}
F. Pi\'echon, A. Raoux, J.-N. Fuchs, and G. Montambaux,
``Geometric orbital susceptibility: Quantum metric without Berry curvature",
Phys. Rev. B \textbf{94}, 134423 (2016).

\bibitem{ozawa18}
T. Ozawa,
``Steady-state Hall response and quantum geometry of driven-dissipative lattices",
Phys. Rev. B \textbf{97}, 041108(R) (2018).


\bibitem{torma15}
S. Peotta and P. T\"{o}rm\"{a}, 
``Superfluidity in topologically nontrivial flat bands",
Nat. Commun. {\bf 6}, 8944 (2015).

\bibitem{torma16}
A. Julku, S. Peotta, T. I. Vanhala, D.-H. Kim, and P. T\"{o}rm\"{a},
``Geometric Origin of Superfluidity in the Lieb-Lattice Flat Band",
Phys. Rev. Lett. \textbf{117}, 045303 (2016).

\bibitem{iskin18}
M. Iskin,
``Berezinskii-Kosterlitz-Thouless transition in the time-reversal-symmetric Hofstadter-Hubbard model",
Phys. Rev. A \textbf{97}, 013618 (2018).

\bibitem{iskin18a}
M. Iskin, 
``Quantum metric contribution to the pair mass in spin-orbit coupled Fermi superfluids",
Phys. Rev. A \textbf{97}, 033625 (2018).
 
\bibitem{iskin17}
M. Iskin, 
``Exposing the quantum geometry of spin-orbit coupled Fermi superfluids", 
Phys. Rev. A \textbf{97}, 063625 (2018).

\bibitem{iskin18b}
M. Iskin, 
``Spin-susceptibility of spin-orbit coupled Fermi superfluids",
Phys. Rev. A \textbf{97}, 053613 (2018).



\bibitem{provost80}
J. P. Provost and G. Vallee,
``Riemannian structure on manifolds of quantum states",
Commun. Math. Phys. \textbf{76}, 289 (1980).

\bibitem{berry89}
M. V. Berry,
``The quantum phase, five years after in Geometric Phases in Physics",
edited by A. Shapere and F. Wilczek (World Scientific, Singapore, 1989).

\bibitem{thouless98}
D. J. Thouless,
``Topological Quantum Numbers in Nonrelativistic Physics",
(World   Scientific,  Singapore,1998)


\bibitem{am76}
N. W. Ashcroft and N. D. Mermin, 
\textit{Solid State Physics}, 
1st ed. (Harcourt College Publishers, 1976).

%

\bibitem{zhai11}
Z.-Q. Yu and H. Zhai,
``Spin-Orbit Coupled Fermi Gases across a Feshbach Resonance",
Phys. Rev. Lett. \textbf{107}, 195305 (2011).

\bibitem{iskin11}
M. Iskin and A. L. Suba\c{s}\i, 
``Quantum phases of atomic Fermi gases with anisotropic spin-orbit coupling", 
Phys. Rev. A \textbf{84}, 043621 (2011).

\bibitem{jiang11}
L. Jiang, X.-J. Liu, H. Hu, and H. Pu,
``Rashba spin-orbit-coupled atomic Fermi gases",
Phys. Rev. A \textbf{84}, 063618 (2011).

\bibitem{hu11}
H. Hu, L. Jiang, X.-J. Liu, and H. Pu,
``Probing Anisotropic Superfluidity in Atomic Fermi Gases with Rashba Spin-Orbit Coupling",
Phys. Rev. Lett. \textbf{107}, 195304 (2011).

\bibitem{he12a}
L. He and X.-G. Huang,
``BCS-BEC Crossover in 2D Fermi Gases with Rashba Spin-Orbit Coupling",
Phys. Rev. Lett. \textbf{108}, 145302 (2012).

\bibitem{he12b}
L. He and X.-G. Huang,
``BCS-BEC crossover in three-dimensional Fermi gases with spherical spin-orbit coupling",
Phys. Rev. B \textbf{86}, 014511 (2012).

\bibitem{shenoy12a}
J. P. Vyasanakere and V. B. Shenoy,
``Rashbons: properties and their significance",
New J. of Phys. \textbf{14}, 043041 (2012).

\bibitem{shenoy12b}
J. P. Vyasanakere and V. B. Shenoy,
``Collective excitations, emergent Galilean invariance, and boson-boson interactions across the BCS-BEC crossover induced by a synthetic Rashba spin-orbit coupling",
Phys. Rev. A \textbf{86}, 053617 (2012).

\bibitem{torma18} Noting that the band minimum of the lower helicity band is 
locally flat for all $\mathbf{k_m}$ states, our results are also consistent with 
that of P. T\"{o}rm\"{a},  L. Liang, and S. Peotta,
``Quantum metric and effective mass of a two-body bound state in a flat band",
Phys. Rev. B \textbf{98}, 220511(R) (2018).
For an energetically isolated flat band in $\mathbf{k}$ space, the inverse mass 
tensor of the two-body bound state is reported as
$
\hbar^2 [\mathbb{M}^{-1}_{tb}]^{ij} = |E_0| \langle g_{\mathbf{k}}^{ij} \rangle,
$
where $E_0 = -U/2$ is the energy of the zero-momentum pair with $U \le 0$ 
characterizing the short-ranged attractive interactions, and 
$
\langle g_{\mathbf{k}}^{ij} \rangle = (1/N) \sum_\mathbf{k} g_{\mathbf{k}}^{ij}
$
is an average over the entire $\mathbf{k}$ space with $N$ degenerate points.
In connection to our case, reexpressing the average effective mass as
$
\hbar^2 \langle [\mathbb{M}^{-1}_{-,\mathbf{k_m}}]^{ij} \rangle 
= \hbar^2 \delta_{ij}/m_0 - 2m_0 \alpha^2 \langle g_{\mathbf{k_m}}^{ij} \rangle,
$
we directly identify the average
$
|E_0| \langle g_{\mathbf{k_m}}^{ij} \rangle
$
over the degenerate subspace as the geometric contribution to the inverse pair 
mass tensor, where $E_0 = -m_0\alpha^2$ is the energy of the zero-momentum 
pair. See~\cite{iskin18a} for a similar observation, and note~\cite{boltzmann}.

\bibitem{boltzmann}
For the isotropic systems of interest in this paper, we define the relevant effective 
mass as the $\mathbf{k}$-space average of Eq.~(\ref{eqn:emt2}) over the 
degenerate subspace. Such an average mass may also be motivated from 
the electrical conductivity tensor $\boldsymbol{\sigma}_{n}^{ij}$. For instance, 
within the semiclassical Boltzmann transport equation, integrating the tensor
$
\boldsymbol{\sigma}_{n}^{ij} \propto \sum_\mathbf{k} 
v_{n \mathbf{k}}^{i} v_{n \mathbf{k}}^{j} 
[-\partial f(\varepsilon_{n\mathbf{k}} - \mu)/\partial \varepsilon]
$
by parts, we may reexpress it as 
$
\boldsymbol{\sigma}_{n}^{ij} = \propto \sum_\mathbf{k}
[\mathbb{M}^{-1}_{n \mathbf{k}}]^{ij} f(\varepsilon_{n\mathbf{k}} - \mu).
$
Here, $f(x) = 1/[e^{x/(k_B T)} + 1]$ is the Fermi-Dirac distribution with $k_B$ 
the Boltzmann constant and $T$ the temperature, and $\mu$ is the chemical 
potential. See page 250 in~\cite{am76} for details.

\end{thebibliography}
\end{document}